\documentclass[fleqn,11pt,twoside]{article}
\usepackage{amsthm,amsthm,amssymb, color, epsfig, graphics, subfigure,}
\usepackage{tikz}

\usepackage{amsmath}
\usepackage{graphicx}
\usepackage{amsmath}
\usepackage{graphicx}
\usepackage{latexsym}
\usepackage[all]{xy}
\usepackage{amsthm}
\usepackage{amsmath, amscd}
\usepackage{graphicx}
\usepackage{lscape}
\usepackage{amsmath}
\usepackage{graphicx}
\usepackage{latexsym}

\newtheorem{remark}{Remark}

\makeatletter
\newcommand{\copyrightnote}[2]{{\renewcommand{\thefootnote}{}
 \footnotetext{\small\it
\begin{flushleft}
 \copyright \ #1   #2
\end{flushleft}}}}

\newcommand{\Name}[1]{\begin{flushleft}
                       \LARGE \bf #1
                       \end{flushleft}\vspace{-3mm}}

\newcommand{\Author}[1]{\begin{flushleft}
                       \it #1 \end{flushleft}}

\newcommand{\Address}[1]{\begin{flushleft}
                       \it #1 \end{flushleft}}

\newcommand{\Date}[1]{\begin{flushleft}
                      \small  \it #1 \end{flushleft}}

\newcommand{\evenhead}{Author \ name}
\newcommand{\oddhead}{Article \ name}

\renewcommand{\@evenhead}{
\hspace*{-3pt}\raisebox{-15pt}[\headheight][0pt]{\vbox{\hbox to \textwidth
{\thepage \hfil \evenhead}\vskip4pt \hrule}}}
\renewcommand{\@oddhead}{
\hspace*{-3pt}\raisebox{-15pt}[\headheight][0pt]{\vbox{\hbox to \textwidth
{\oddhead \hfil \thepage}\vskip4pt\hrule}}}
\renewcommand{\@evenfoot}{}
\renewcommand{\@oddfoot}{}

\long\def\@makecaption#1#2{%
  \vskip\abovecaptionskip
  \sbox\@tempboxa{\small \textbf{#1.}\ \ #2}%
  \ifdim \wd\@tempboxa >\hsize
    {\small \textbf{#1.}\ \ #2}\par
  \else
    \global \@minipagefalse
    \hb@xt@\hsize{\hfil\box\@tempboxa\hfil}%
  \fi
  \vskip\belowcaptionskip}

\newcommand{\JNMPnumberwithin}[3][\arabic]{%
  \@ifundefined{c@#2}{\@nocounterr{#2}}{%
    \@ifundefined{c@#3}{\@nocnterr{#3}}{%
      \@addtoreset{#2}{#3}%
      \@xp\xdef\csname the#2\endcsname{%
        \@xp\@nx\csname the#3\endcsname .\@nx#1{#2}}}}%
}

\newcommand{\resetfootnoterule} {
  \renewcommand\footnoterule{%
  \kern-3\p@
  \hrule\@width.4\columnwidth
  \kern2.6\p@}
}

\renewcommand{\footnoterule}{}

\newcommand{\be}{\begin{equation}}
\newcommand{\ee}{\end{equation}}
\newcommand{\ba}{\hspace*{-5pt}\begin{array}}
\newcommand{\ea}{\end{array}}
\newcommand{\p}{\partial}

\makeatother

\numberwithin{equation}{section}
\theoremstyle{definition}

\renewcommand{\ba}{\begin{array}}
\renewcommand{\ea}{\end{array}}
\newcommand{\beg}{\begin{eqnarray}}
\newcommand{\eeq}{\end{eqnarray}}
\newcommand{\bg}{\begin{eqnarray*}}

\newcommand{\ed}{\end{eqnarray*}}

\newcommand{\nn}{\nonumber}

\renewcommand{\p}{\partial} 
 
\newcommand{\notlhd}{\lhd\kern-.8em{/}\ } 
\newcommand{\notexist}{\ \exists\kern-.5em{\raise.1em\hbox{/}}\ }

\newcommand{\pde}[2]{\frac{\p #1}{\p #2}}

\newcommand{\inp}{{\mbox{\vbox{\hrule width0ex\hbox{\vrule
 height0ex\kern3.8pt
\vbox{\kern2.5pt}\kern3.8pt \vrule height1.6ex}
\hrule width1.6ex}}}}

\begin{document}


\renewcommand{\evenhead}{ {\LARGE\textcolor{blue!10!black!40!green}{{\sf \ \ \ ]ocnmp[}}}\strut\hfill M. Euler, N. Euler and M.C. Nucci}
\renewcommand{\oddhead}{ {\LARGE\textcolor{blue!10!black!40!green}{{\sf ]ocnmp[}}}\ \ \ \ \ Differential equations and the 2-variable Möbius transformations}


\thispagestyle{empty}

\newcommand{\FistPageHead}[3]{
\begin{flushleft}
\raisebox{8mm}[0pt][0pt]
{\footnotesize \sf
\parbox{150mm}{{Open Communications in Nonlinear Mathematical Physics}\ \ \ {\LARGE\textcolor{blue!10!black!40!green}{]ocnmp[}}
\ \ Vol.2 (2022) pp
#2\hfill {\sc #3}}}\vspace{-13mm}
\end{flushleft}}

\FistPageHead{1}{\pageref{firstpage}--\pageref{lastpage}}{ \ \ Letter}

\strut\hfill

\strut\hfill

\copyrightnote{The author(s). Distributed under a Creative Commons Attribution 4.0 International License}

\qquad\qquad\qquad\qquad\qquad\qquad {\LARGE  {\sf Letter to the Editors}}

\strut\hfill

%
%







%

\Name{On differential equations invariant under two-variable Möbius transformations}

\label{firstpage}




\Author{M Euler$^{\,1}$, N Euler$^{\,1\, *}$ and M C Nucci$^{\,2}$  }



\Address{
$^1$ Centro Internacional de Ciencias, Av. Universidad s/n, Colonia Chamilpa,\\
 62210 Cuernavaca, Morelos, Mexico\\[0.3cm]
$^2$ Department of Mathematics, University of Bologna, Piazza di Porta San Donato 5, Bologna, Italy\\[0.3cm]
${\ }^*$ Corresponding author's email address: Dr.Norbert.Euler@gmail.com}

\Date{Received October 25, 2022; Accepted November 29, 2022}





\noindent
{\bf Abstract}: 
We compute invariants for the two-variable Möbius transformation. In particular we are interested in partial differential equations in two dependent and two independent variables that are kept invariant under this transformation.

\strut\hfill



\renewcommand{\theequation}{\arabic{section}.\arabic{equation}}

\allowdisplaybreaks

\section{Introduction: motivation from the one-variable Möbius transformation}

In this section we discuss in some detail the invariants of the one-variable Möbius transformation
\begin{gather}
\label{Mobius-u}  
{\cal M}: 
\displaystyle{
u\mapsto v= \frac{\alpha_1 u+\beta_1}{\alpha_2 u+\beta_2}},
\end{gather}
for two cases, namely $u=u(x)$ and $u=u(x,t)$. 
This serves as a motivation for the current study of the two-variable Möbius transformation for two dependent variables $\{u_1,\,u_2\}$ in two independent variables $\{x,\,t\}$. In fact, we have reported 
results for the one independent variable case in 
\cite{Euler-Euler-Nucci-AML17}, and the current paper is a continuation of that study.
In particular, we note the importance of the {\it Schwarzian derivative} $S$ (see (\ref{Schwarzian})) which is an invariant of the Möbius transformation 
(\ref{Mobius-u}). This is well known, but we use the opportunity to discuss this here again in order to compare the results with the case of the
two-variable Möbius transformation that is introduced in Section 2 (see transformation (\ref{M-u-n})). 
Instead of the 3rd-order invariant for the one-variable Möbius transformation, namely the Schwarzian derivative (\ref{Schwarzian}), we obtain two 4th-order invariants for the two-variable Möbius transformation.
We consider this an interesting finding that could be of importance in the study of integrable systems of nonlinear ordinary and partial differential equations.

We recall that, under the condition $\alpha_1\beta_2-\alpha_2\beta_1=1$, the Möbius transformation
(\ref{Mobius-u}) is associated with the Special Linear 
Transformation $SL(2,\mathbb R)$, where
\begin{gather}
\Phi=\left(
\ba{cc}
\alpha_1&\beta_1\\
\alpha_2&\beta_2
\ea
\right)
\in SL(2,\mathbb R),\qquad \mbox{det} \Phi=1.
\end{gather}
A basis for the corresponding 3-dimensional Lie algebra $sl(2,\mathbb R)$ is
given by the three matrices
\begin{gather}
X_1=
\left(
\ba{cc}
0&1\\
0&0
\ea
\right),
\qquad
X_2=
\left(
\ba{cr}
1&0\\
0&-1
\ea
\right),
\qquad
X_3=
\left(
\ba{cr}
0&0\\
1&0
\ea
\right).
\end{gather}
We recall that
\begin{gather}
\Phi=\exp(\epsilon X),\quad\mbox{where}\ \Phi\in\mbox{SL$(2,\mathbb R)$ and}\ X\in\mbox{sl$(2,\mathbb R)$},
\end{gather}
where $\epsilon$ is a small real parameter.
Since $\det \Phi=\exp(\epsilon\, \mbox{Tr}\,X)$ and $\det \Phi=1$,
we have Tr$\,(X_j)=0$ for all $j=1,2,3$. For a given infinitesimal transformation
\begin{gather}
u\mapsto \varphi(u;\epsilon)
\end{gather}
the Lie generator takes the form 
\begin{gather}
\left. 
Z=\left(\pde{\varphi(u;\epsilon)}{\epsilon}
\right|_{\epsilon=0}\right) \pde{\ }{u}.
\end{gather}
This leads to the following set of basis Lie generators for $sl(2,\mathbb R)$:
\begin{gather}
\label{Z-u}
\{\pde{\ }{u},\ u\pde{\ }{u},\ u^2\pde{\ }{u}\}.
\end{gather}
For more details we refer to \cite{Olver-book} (see also the classical work \cite{Classical})

In order to compute the invariants $I$ of order $p$ for the transformation (\ref{Mobius-u}) one needs to solve the linear system of partial differential equations given by the condition
\begin{gather}
Z_j^{(p)} I=0,\quad j=1,2,3.
\end{gather}
Here $Z_j^{(p)}$ denotes the prolongation of the Lie generator $Z_j$ up to order $p$ (see \cite{Olver-book} for details).

\subsection{One dependent variable and one independent variable: $u(x)$}
The simplest case is given by the mapping of one dependent variable $u$ for one independent variable $x$, i.e. $u=u(x)$. 
For the Möbius transformation (\ref{Mobius-u}), we now have 
\begin{gather}
\label{Mobius-u-one-x}  
{\cal M}: 
\left\{
\ba{l}
\displaystyle{
u({x})\mapsto v(\bar{x})= \frac{\alpha_1 u({x})+\beta_1}{\alpha_2 u({x})+\beta_2}}\\
\\
\displaystyle{{x}\mapsto \bar {x}={x}}\\
\ea
\right.
\end{gather}
where
\begin{gather}
\det\left(
\ba{cc}
\alpha_1&\beta_1\\
\alpha_2&\beta_2
\ea
\right)=1.
\end{gather}
Computing the invariants for this transformation up to order 3, we obtain 
\begin{subequations}
\begin{gather}
\omega_0=f(x)\\
\label{Schwarzian}
S=\frac{u_{xxx}}{u_x}
-\frac{3}{2}\frac{u_{xx}^2}{u_x^2},
\end{gather}
\end{subequations}
where $f$ is an arbitrary smooth function and $S$ is the well-know Schwarzian derivative \cite{Ovsienko}. In fact, all higher-order invariants consist of 
$x$-derivatives of $S$. We can now state that any $n$th-order ordinary differential equation of the form
\begin{gather}
\Psi(x,S,S_x,S_{xx},\ldots ,S_{nx})=0
\end{gather}
is invariant under the given Möbius transformation for arbitrary smooth $\Psi$. Here $S_{nx}$ denotes the $n$th derivative of $S$ with respect to $x$.

\strut\hfill

\noindent
{\bf Example 1.1:} We consider
\begin{gather}
S_x=\Psi(x,S),
\end{gather}
which is the following 4th-order equation:
\begin{gather}
\label{ODE-example-4th-order}
u_{4x}=\frac{4u_{xxx}u_{xx}}{u_x}
-\frac{3u_{xx}^3}{u_x^2}+u_x\Psi(x,S).
\end{gather}
Equation (\ref{ODE-example-4th-order}) is clearly invariant under the current Möbius transformation (\ref{Mobius-u-one-x}) and it
 is moreover the most general semilinear ordinary differential equation of order four that admits the Lie symmetry algebra $sl(2,\mathbb R)$ with basis (\ref{Z-u}).
Equation (\ref{ODE-example-4th-order}) with $\Psi=\Psi(S)$ has been reported in \cite{Nucci-JNMP-2002} for the 4-dimensional Lie symmetry algebra 
\begin{gather}
\{\pde{\ }{x}, \pde{\ }{u},u\pde{\ }{u},u^2\pde{\ }{u}\}.
\end{gather}

\subsection{One dependent variable and two independent variables: $u(x,t)$}

We compute all invariants $I$, up to order three, for the Möbius transformation that maps
$u(x,t)$. That is
\begin{gather}
\label{Mobius-u-two-x-t}  
{\cal M}: 
\left\{
\ba{l}
\displaystyle{
u({x,t})\mapsto v(\bar{x},\bar{t})= \frac{\alpha_1 u({x,t})+\beta_1}{\alpha_2 u({x,t})+\beta_2}}\\
\\
\displaystyle{{x}\mapsto \bar {x}={x}}\\
\\
\displaystyle{{t}\mapsto \bar {t}={t}},\\
\ea
\right.
\end{gather}
where
\begin{gather}
\det\left(
\ba{cc}
\alpha_1&\beta_1\\
\alpha_2&\beta_2
\ea
\right)=1.
\end{gather}

\noindent
The general solution of 
\begin{gather}
\label{cond-12}
Z_j^{(3)}I=0,  \quad j=1,2,3,
\end{gather}
is
\begin{gather}
I=\Psi(\omega_0,  \omega_1,\omega_2,\omega_3,\omega_4,\omega_5,\omega_6,\omega_7),
\end{gather}
whereby $\omega_0=f(x,t)$ is a trivial invariant and $\omega_k$ are as follows:
\begin{subequations}
\begin{gather}
\label{sl2-inv-1}
\omega_1=\frac{u_t}{u_x}
\\[0.3cm]
\label{sl2-inv-2}
\omega_2=
\omega_{1,x}\\[0.3cm]
\label{sl2-inv-3}
\omega_3
=\omega_{1,t}+\omega_1\omega_2\\[0.3cm]
\label{sl2-inv-4}
\omega_4\equiv S=\frac{u_{xxx}}{u_x}
-\frac{3}{2}\frac{u_{xx}^2}{u_x^2}:\ \mbox{the Schwarzian derivative}
\\[0.3cm]
\label{sl2-inv-5}
\omega_5=\frac{u_{xxt}}{u_x}
-\frac{2u_{xx}u_{xt}}{u_x^2}
+\frac{1}{2}\frac{u_tu_{xx}^2}{u_x^3}=\omega_{2,x}+\omega_1\omega_4\\[0.3cm]
\omega_6=\frac{u_{xtt}}{u_x}
\label{sl2-inv-6}
-\frac{u_{xx}u_{tt}}{u_x^2}
-\frac{2u_tu_{xx}u_{xt}}{u_x^3}
+\frac{3}{2}\frac{u_t^2u_{xx}}{u_x^4}=\omega_{3,x}+\omega_1^2\omega_4\\[0.3cm]
\label{sl2-inv-7}
\omega_7=\frac{u_{ttt}}{u_x}
-\frac{3u_tu_{xx}u_{tt}}{u_x^3}
+\frac{3}{2}\frac{u_t^3u_{xx}^2}{u_x^5}=\omega_{3,t}+\omega_2\omega_3+\omega_1^3\omega_4.
%
\end{gather}
\end{subequations}

\noindent
From the set of invariants, $\{\omega_0,\omega_1,\omega_2,\ldots,\omega_7\}$ we note that two stand out, namely $\omega_1$ and the Schwarzian derivative 
$\omega_4$, which are fundamental for this transformation in the sense that all the other invariants can be expressed as functions of $\omega_1$, $\omega_4$ and its derivatives. 


\strut\hfill

\noindent
Using the two invariants (\ref{sl2-inv-1}) and (\ref{sl2-inv-4}), as well as their $x$ and $t$-derivatives of any order, one can write partial differential equations of the form  
\begin{gather}
\label{PDE-3rd-gen-prop2}
\Psi(\omega_0,\omega_1, S, \omega_{1,x},\omega_{1,t}, \omega_{1,xx}, \ldots, S_x,S_{t},S_{xx},\ldots)=0,
\end{gather}
all of which are invariant under the Möbius transformation (\ref{Mobius-u-two-x-t}) and all of which admit (\ref{Z-u}) as its Lie symmetry algebra. 
Of particular interest are the autonomous evolution equations
\begin{gather}
\label{Evolution-Eq-M-Inv-Gen}
u_t=u_x\Psi(S,S_x,\ldots,S_{(m-3)x}),
\end{gather}
where $m\geq 3$. We discuss this in Example 1.2.

\strut\hfill

\noindent
{\bf Example 1.2:}
An interesting question regarding nonlinear partial differential equations of the class 
(\ref{Evolution-Eq-M-Inv-Gen})
to find those equations that are {\it symmetry-integrable}. A symmetry-integrable equation is defined as an equations that admits an infinite set of local generalized (or Lie-Bäcklund) symmetries.
In \cite{E-E-April2019} we have addressed this problem and consequently obtained a class of symmetry-integrable
3rd-order and 5th-order autonomous evolution equations of the form (\ref{Evolution-Eq-M-Inv-Gen}).
In particular, the 3rd-order evolution equations of the form  
\begin{gather}
\label{3rd-order-Evolution-general}
u_t=u_x\Psi(S),
\end{gather}
contains the following class of equations that are symmetry-integrable \cite{E-E-April2019}:
\begin{subequations}
\begin{gather}
\label{SKdV}
u_t=u_xS\ :\quad \mbox{the Schwarzian KdV equation}\\[0.3cm]
\label{full-non-1-intro}
u_t=-2\frac{u_x}{\sqrt{S}}\\[0.3cm]
\label{full-non-2-intro}
u_t=\frac{u_x}{(b_1-S)^2}\\[0.3cm]
\label{full-non-3-intro}
u_t=\frac{u_x}{S^2}\\[0.3cm]
\label{full-non-4-intro}
u_t=u_x\left( \frac{a_1-S}{(a_1^2+3a_2)(S^2-2a_1S-3a_2)^{1/2}}\right).
\end{gather}
\end{subequations}
Here $S$ is the Schwarzian derivative (\ref{Schwarzian}) and the constants $a_1,\ a_2$ and $b_1$ are arbitrary, except for the conditions
 $a_1^2+3a_2\neq 0$ and $b_1\neq 0$. 
 In \cite{E-E-JNMP2021} we propose a method by which one can compute higher-order symmetry-integrable and Möbius-invariant evolution equations from the above fully-nonlinear third-order equations (\ref{full-non-1-intro}) -- (\ref{full-non-4-intro}).

For the 5th-order semi-linear Möbius-invariant autonomous evolution equations 
\begin{gather}
\label{Evolution eq-1}
u_t=u_x\Psi(S,S_x,S_{xx}),
\end{gather}
the following were found to be symmetry-integrable \cite{E-E-April2019}:
\begin{subequations}
\begin{gather}
\label{SKuperI-intro}
u_t=u_x\left(S_{xx}+\frac{1}{4}S^2\right): \quad \mbox{the Schwarzian Kupershmidt I equation;}\\[0.3cm]
\label{SKuperII-intro}
u_t=u_x\left(S_{xx}+4S^2\right): \quad \mbox{the Schwarzian Kupershmidt II equation;}\\[0.3cm]
\label{SKdV-5thOrder-intro}
u_t=u_x\left(S_{xx}+\frac{3}{2}S^2\right)\quad \mbox{the Schwarzian 5th-order KdV equation.}
\end{gather}
\end{subequations}
 The recursion operators that generate the higher-order members in the symmetry-integrable hierarchies of (\ref{SKuperI-intro}), (\ref{SKuperII-intro}) and  (\ref{SKdV-5thOrder-intro}) are given in \cite{E-E-April2019}. The three Möbius-invariant equations
  (\ref{SKuperI-intro}), (\ref{SKuperII-intro}) and  (\ref{SKdV-5thOrder-intro}) play a central role for a large class of 5th-order symmetry-integrable equations that are related to these equations in terms of nonlocal transformations by the so-called {\it multipotentialization procedure} (see \cite{Euler-Euler-book-article} for details).

\strut\hfill

\noindent
{\remark
Replacing $t$ by $x$ for the invariants (\ref{sl2-inv-5}), (\ref{sl2-inv-6}) and (\ref{sl2-inv-5}), results in the Schwarzian derivative (\ref {sl2-inv-4}). One could therefore think of 
 (\ref{sl2-inv-5}), (\ref{sl2-inv-6}) and (\ref{sl2-inv-5}) as two-dimensional Schwarzian-type derivatives involving two independent variables $x$ and $t$. Let us denote this two-dimensional Schwarzian-type derivative defined by 
 (\ref{sl2-inv-5}), by the notation $S_1^{[x,y]}$, i.e.
 \begin{gather}
 S_1^{[x,y]}u(x,t):=\frac{u_{xxt}}{u_x}-\frac{2u_{xx}u_{xt}}{u_x^2}+\frac{1}{2}\frac{u_tu_{xx}^2}{u_x^3}.
 \end{gather}
 Consider now a function $f(u(x,t))$. Using $S_1^{[x,y]}$ we have
 \begin{gather}
(S_1^{[x,y]}f)(u(x,t))=\frac{D_{xxt}f}{D_xf}
-2\frac{(D_{xx}f)(D_{xt} f)}{(D_xf)^2}
+\frac{1}{2}\frac{(D_tf)(D_{xx}f)^2}{(D_xf)^3}\nn\\[0.3cm]
\qquad\qquad\qquad\quad\ 
=\left(\frac{f_{uuu}}{f_u}-\frac{3}{2}\frac{f_{uu}^2}{f_u^2}\right)u_xu_t
+S_1^{[x,y]}u(x,t),
\end{gather}
where $D$ denote the total derivative and the subscript $u$ the ordinary derivative with respect to $u$. Additional two-dimensional Schwarzian-type derivatives can be defined using the remaining two third-order invariants, that is $S_2^{[x,y]}:=\omega_6$ and $S_3^{[x,y]}:=\omega_7$, 
where $\omega_6$ and $\omega_7$ are given by (\ref{sl2-inv-6}) and (\ref{sl2-inv-5}), respectively. One should remark that the invariants listed above are of course not unique, as any function of
the given invariants is an invariant for the same transformation and, moreover, the roles of $x$ and $t$ can be exchanged in the derivatives of the expressions  (\ref{sl2-inv-1}) to (\ref{sl2-inv-7}) without affecting the invariance under the Möbius transformation (\ref{Mobius-u-two-x-t}). 
}

\strut\hfill 

\noindent
{\remark
It is interesting to note that the Bateman equation
\begin{gather}
\label{bateman}
u_{tt}u_x^2-2u_xu_tu_{xt}+u_{xx}u_t^2=0
\end{gather}
can be presented in terms of the invariants (\ref{sl2-inv-1}), (\ref{sl2-inv-2}) and (\ref{sl2-inv-3}), namely as
follows:
\begin{gather}
(\omega_3-2\omega_1\omega_2)u_x^3=0.
\end{gather}
Equation (\ref{bateman}) plays an important role in the Painlev\'e analysis of PDEs (see e.g. \cite{Steeb-Euler-P-book} and \cite{Euler-Lindblom-2000}). It is clear that 
$u_{tt}u_x^2-2u_xu_tu_{xt}+u_{xx}u_t^2$ is not invariant 
under the Möbius transformation (\ref{Mobius-u-two-x-t}). However, (\ref{bateman}) does admit the Lie symmetry 
algebra $sl(2,\mathbb R)$ with basis generators (\ref{Z-u}).  This can directly be verified by the Lie symmetry condition
\begin{gather}
\left.
\vphantom{\frac{DA}{DB}}
Z_j^{(2)} E\right|_{E=0}=0,\ j=1,2,3,
\end{gather}
where $E:=u_{tt}u_x^2-2u_xu_tu_{xt}+u_{xx}u_t^2$ and $\{Z^{(2)}_1,  Z^{(2)}_2,Z^{(2)}_3\}$ are the 2nd-prolongations of the generators (\ref{Z-u}).
}

\section{The Möbius transformation that maps two variables}

We now consider the Möbius transformation 
\begin{gather}
\label{M-u-n}
{\cal M}: 
\displaystyle{{\bf u}\mapsto {\bf v}=\frac{A{\bf u}+B}{C{\bf u}+\beta}},
\end{gather}
where ${\bf u}=(u_1,u_2)$, 
${\bf v} =(v_1 ,v_2)$, 
$A$ is a constant $2\times 2$ matrix, 
\begin{gather}
A=\left(
\ba{cc}
a_{11}&a_{12}\\
a_{21}&a_{22}
\ea
\right),
\end{gather}
$B$ is a constant $2\times 1$ matrix and $C$ a constant $1\times 2$ matrix,
\begin{gather}
B=\left(
\ba{c}
b_{11}\\
b_{21}
\ea
\right),\qquad
C=\left(
\ba{cccc}
c_{11}&
c_{12}
\ea
\right),
\end{gather}
and $\beta$ is a constant, with the condition 
\begin{gather}
\mbox {Det}\,
\left(
\ba{cc}
A&B\\
C&\beta
\ea
\right)=1.
\end{gather}
Now
\begin{gather}
\Phi=\left(
\ba{cc}
A&B\\
C&\beta
\ea
\right)\in SL(3,\mathbb R).
\end{gather}
and $sl(3,\mathbb R)$ is the Lie algebra of all real traceless $3 \times 3$ matrices
with dim$[sl(3,\mathbb R)]=8$.
A basis for the 8-dimensional Lie algebra $sl(3,\mathbb R)$ is
given by the following eight matrices:
\begin{gather}
X_1=
\left(
\ba{ccc}
0&0&1\\
0&0&0\\
0&0&0
\ea
\right),
\qquad\ \ 
X_2=
\left(
\ba{ccc}
0&0&0\\
0&0&1\\
0&0&0
\ea
\right),
\qquad
X_3=
\left(
\ba{crc}
1&0&0\\
0&-1&0\\
0&0&0
\ea
\right),\nonumber\\[0.3cm]
X_4=
\left(
\ba{ccr}
0&0&0\\
0&1&0\\
0&0&-1
\ea
\right),
\qquad
X_5=
\left(
\ba{ccc}
0&0&0\\
1&0&0\\
0&0&0
\ea
\right),
\qquad
X_6=
\left(
\ba{ccc}
0&1&0\\
0&0&0\\
0&0&0
\ea
\right),\\[0.3cm]
X_7=
\left(
\ba{ccc}
0&0&0\\
0&0&0\\
1&0&0
\ea
\right),\qquad\ \ 
X_8=
\left(
\ba{ccc}
0&0&0\\
0&0&0\\
0&1&0
\ea
\right).\nn
\end{gather}
For the infinitesimal transformation
\begin{subequations}
\begin{gather}
v_1=\varphi_1(u_1,u_2;\epsilon)\\
v_2=\varphi_2(u_1,u_2;\epsilon)
\end{gather}
\end{subequations}
the Lie generator takes the form 
\begin{gather}
\left. \left.
Z=\left(\pde{\varphi_1(u_1,u_2;\epsilon)}{\epsilon}
\right|_{\epsilon=0}\right) \pde{\ }{u_1}
+\left(\pde{\varphi_2(u_1,u_2;\epsilon)}{\epsilon}
\right|_{\epsilon=0}\right) \pde{\ }{u_2}.
\end{gather}
This leads to the following basis set of Lie generators for $sl(3,\mathbb R)$:
\begin{gather}
\{Z_1=\pde{\ }{u_1},\ Z_2=\pde{\ }{u_2},\ Z_3=u_1\pde{\ }{u_1},\ Z_4=u_2\pde{\ }{u_2},\
Z_5=u_1\pde{\ }{u_2},\ Z_6=u_2\pde{\ }{u_1},\nonumber\\[0.3cm]
\label{basis-sl3}
\ \ 
Z_7=u_1^2\pde{\ }{u_1}+u_1u_2\pde{\ }{u_2},\ 
Z_8=u_1u_2\pde{\ }{u_1}+u_2^2\pde{\ }{u_2}
\}.
\end{gather}

\subsection{Two dependent and one independent variables: $\{u_1(x),\ u_2(x)\}$}

We consider the Möbius transformation (\ref{M-u-n}) for 
${\bf u}(x)=(u_1(x),u_2(x))$, that is 
\begin{gather}
\label{M-u-2-x}
{\cal M}: 
\left\{
\ba{l}
\displaystyle{{u_1}({x})\mapsto {v_1}(\bar{x})=\frac{a_{11}u_1(x)+a_{12}u_2(x)+b_{11}}{c_{11}u_1(x)+c_{12}u_2(x)+\beta}    }\\
\\
\displaystyle{{u_2}({x})\mapsto {v_2}(\bar{x})=\frac{a_{21}u_1(x)+a_{22}u_2(x)+b_{21}}{c_{11}u_1(x)+c_{12}u_2(x)+\beta}    }\\
\\
\displaystyle{{x}\mapsto \bar {x}={x}}\\
\ea
\right.
\end{gather}
where
\begin{gather}
\mbox {Det}\,
\left(
\ba{ccc}
a_{11}&a_{12}&b_{11}\\
a_{21}&a_{22}&b_{21}\\
c_{11}&c_{12}&\beta
\ea
\right)=1.
\end{gather}
This case has already been studied and is reported in \cite{Euler-Euler-Nucci-AML17}. 
Besides the obvious invariant $\omega_0=f(x)$, (\ref{M-u-2-x}) admits the following
two invariants of order four:
\begin{subequations}
\begin{gather}
\omega_{41}=
-\frac{3u_{1,x}u_{2,x}(S_{1,x}-S_{2,x})}{u_{1,x}u_{2,xx}-u_{2,x}u_{1,xx}}
-4\left(
\frac{u_{1,x}u_{2,x}}{u_{1,x}u_{2,xx}-u_{2,x}u_{1,xx}}\right)^2
(S_1-S_2)^2
\nn\\[0.3cm]
\label{old-w1}
\qquad
+6\left(
\frac{u_{1,x}u_{2,xx}S_1-u_{2,x}u_{1,xx}S_2}{u_{1,x}u_{2,xx}-u_{2,x}u_{1,xx}}\right)\\[0.3cm]
\omega_{42}=
\frac{u_{1,x}u_{2,xx}S_{2,x}-u_{2,x}u_{1,xx}S_{1,x}}{
u_{1,x}u_{2,xx}-u_{2,x}u_{1,xx}}\nn\\[0.3cm]
\qquad
+\frac{2}{3}
\left(\frac{u_{1,x}u_{2,x}}
{u_{1,x}u_{2,xx}-u_{2,x}u_{1,xx}}\right)^2
(S_{1,x}-S_{2,x})(S_1-S_2)\nn\\[0.3cm]
\qquad
+\frac{2}{9}
\left(\frac{3u_{1,xx}(S_1-S_2)}{u_{1,x}u_{2,xx}-u_{2,x}u_{1,xx}}
-\frac{2u_{2,x}u_{1,x}^2(S_1-S_2)^2}{(u_{1,x}u_{2,xx}-u_{2,x}u_{1,xx})^2}\right)\times\nn\\[0.3cm]
\label{old-w2}
\qquad\quad
\times\left(
3u_{2,xx}-\frac{2u_{1,x}u_{2,x}^2(S_1-S_2)}{u_{1,x}u_{2,xx}-u_{2,x}u_{1,xx}}\right).
\end{gather}
\end{subequations}
Here $S_1$ and $S_2$ denote the Schwarzian derivatives for $u_1(x)$ and $u_2(x)$, respectively. That is
\begin{gather}
\label{Schwarzian-u1-u2}
S_1=\frac{u_{1,xxx}}{u_{1,x}}
-\frac{3}{2}\frac{u_{1,xx}^2}{u_{1,x}^2},\quad
S_2=\frac{u_{2,xxx}}{u_{2,x}}
-\frac{3}{2}\frac{u_{2,xx}^2}{u_{2,x}^2}.
\end{gather}
Systems of ordinary differential equations invariant under (\ref{M-u-2-x}) and related to
the above invariants (\ref{old-w1}) and (\ref{old-w2}) have been described and reported in \cite{Euler-Euler-Nucci-AML17}.

\subsection{Two dependent and two independent variables: $\{u_1(x,t),\ u_2(x,t)\}$}

Here we consider the Möbius transformation (\ref{M-u-n}) for 
${\bf u}(x,t)=(u_1(x,t),u_2(x,t))$, that is
\begin{gather}
\label{M-u-2-xt}
{\cal M}: 
\left\{
\ba{l}
\displaystyle{{u_1}({x,t})\mapsto {v_1}(\bar{x},\bar{t})=\frac{a_{11}u_1(x,t)+a_{12}u_2(x,t)+b_{11}}{c_{11}u_1(x,t)+c_{12}u_2(x,t)+\beta}    }\\
\\
\displaystyle{{u_2}({x,t})\mapsto {v_2}(\bar{x},\bar{t})=\frac{a_{21}u_1(x,t)+a_{22}u_2(x,t)+b_{21}}{c_{11}u_1(x,t)+c_{12}u_2(x,t)+\beta}    }\\
\\
\displaystyle{{x}\mapsto \bar {x}={x}}\\
\\
\displaystyle{{t}\mapsto \bar {t}={t}},\\
\ea
\right.
\end{gather}
where
\begin{gather}
\mbox {Det}\,
\left(
\ba{ccc}
a_{11}&a_{12}&b_{11}\\
a_{21}&a_{22}&b_{21}\\
c_{11}&c_{12}&\beta
\ea
\right)=1.
\end{gather}
We compute all 3rd-order invariants for (\ref{M-u-2-xt}). For the general solution of
\begin{gather}
\label{Cond-216}
Z_j^{(3)}I=0,\qquad j=1,2,\ldots,8,
\end{gather}
we obtain
\begin{gather}
I=F(\omega_0,\omega_1,\omega_2,\ldots,\omega_{12}),
\end{gather}
where $\omega_0=f(x,t)$ is the trivial invariant, and 
\begin{subequations}
\begin{gather}
\label{sl3-omega-1}
\omega_1=\frac{u_{2,x}u_{1,xx}-u_{1,x}u_{2,xx}}{u_{2,x}u_{1,t}-u_{1,x}u_{2,t}}\\[0.3cm]
\label{sl3-omega-2}
\omega_2=\frac{
u_{1,xx}u_{2,t}
-u_{1,t}u_{2,xx}
+2u_{2,x}u_{1,xt}
-2u_{2,xt}u_{1,x}}
{2u_{2,x}u_{1,t}-2u_{1,x}u_{2,t}}\\[0.3cm]
\label{sl3-omega-3}
\omega_3=\frac{
u_{1,tt}u_{2,t}-u_{1,t}u_{2,tt}}
{u_{2,x}u_{1,t}-u_{1,x}u_{2,t}}\\[0.3cm]
\label{sl3-omega-4}
\omega_4=\frac{
u_{2,x}u_{1,tt}
+2u_{1,xt}u_{2,t}
-2u_{1,t}u_{2,xt}
-u_{2,tt}u_{1,x}}
{u_{2,x}u_{1,t}-u_{1,x}u_{2,t}}
\\[0.3cm]
\label{sl3-omega-5}
\omega_5=2(\omega_{2,x}-\omega_{1,t}-\omega_1\omega_4+\omega_2^2)
\\[0.3cm]
\label{sl3-omega-6}
\omega_6=\omega_{1,x}-2\omega_1\omega_2
\\[0.3cm]
\label{sl3-omega-7}
\omega_7=\frac{2}{3}\omega_{4,x}-\frac{2}{3}\omega_{2,t}-\omega_1\omega_3
\\[0.3cm]
\label{sl3-omega-8}
\omega_8=\omega_{1,t}+\omega_1\omega_4-2\omega_2^2
\\[0.3cm]
\label{sl3-omega-9}
\omega_9=\omega_{3,x}
\\[0.3cm]
\label{sl3-omega-10}
\omega_{10}=-\frac{1}{3}\omega_{4,x}+\frac{4}{3}\omega_{2,t}+2\omega_3
\\[0.3cm]
\label{sl3-omega-11}
\omega_{11}=\omega_{3,t}+\omega_3\omega_4
\\[0.3cm]
\label{sl3-omega-12}
\omega_{12}=-2\omega_{3,x}+\omega_{4,t}+4\omega_2\omega_3+\omega_4^2
\end{gather}
\end{subequations}
Note that there exist no fundamental invariants of order three since all the 3rd-order invariants are combinations of the 2nd-order invariants and its derivatives. 
In addition to the above listed invariants, we recall the 4th-order invariants (\ref{old-w1}) and (\ref{old-w2}) that was obtained for the Möbius transformation (\ref{M-u-2-x}). 
These 4th-order invariants are of course also valid for the Möbius transformation (\ref{M-u-2-xt}), whereby $u_1=u_1(x,t)$ and $u_2=u_2(x,t)$ in
(\ref{old-w1}) and (\ref{old-w2}). All these invariants can easily be verified by checking condition (\ref{Cond-216}).

\strut\hfill

\noindent
Using now the six invariants (\ref{sl3-omega-1}) - (\ref{sl3-omega-4}), (\ref{old-w1}) and (\ref{old-w2}), as well as $x$ and $t$-derivatives of any order of those invariants, one can write systems of partial differential equations,
all of which are invariant under the Möbius transformation (\ref{M-u-2-xt}) and all of which admit (\ref{basis-sl3}) as its Lie symmetry algebra. 

\strut\hfill

\noindent
{\bf Example 2.1:} A system of two partial differential equations of order two that is invariant under the Möbius transformation (\ref{M-u-2-xt}) is the following:
\begin{subequations}
\begin{gather}
u_{1,xx}=
\frac{1}{u_{2,x}u_{1,t}-u_{1,x}u_{2,t}}
\left[
\vphantom{\frac{DA}{DB}}
2u_{1,x}u_{2,x}u_{1,xt}
-2u_{1,x}^2u_{2,xt}\right.\nonumber\\[0.3cm]
\qquad
\left.
\vphantom{\frac{DA}{DB}}
+(u_{2,x}u_{1,t}^2-u_{1,x}u_{1,t}u_{2,t})F_1(\omega_3,\omega_4)
+     2(u_{1,x}^2u_{2,t}-u_{1,x}u_{1,t}u_{2,x})F_2(\omega_3,\omega_4)
\right]\\[0.3cm]
u_{2,xx}=
\frac{1}{u_{2,x}u_{1,t}-u_{1,x}u_{2,t}}
\left[
\vphantom{\frac{DA}{DB}}
-2u_{1,x}u_{2,x}u_{2,xt}
+2u_{2,x}^2u_{1,xt}\right.\nonumber\\[0.3cm]
\qquad
\left.
\vphantom{\frac{DA}{DB}}
-(u_{1,x}u_{2,t}^2-u_{2,x}u_{1,t}u_{2,t})F_1(\omega_3,\omega_4)
-2(u_{2,x}^2u_{1,t}-u_{1,x}u_{2,t}u_{2,x})F_2(\omega_3,\omega_4).
\right],
\end{gather}
\end{subequations}
This is obtained by algebraically solving $u_{1,xx}$ and $u_{2,xx}$ from the system
\begin{subequations}
\begin{gather*}
\omega_1=F_1(\omega_3,\omega_4)\\
\omega_2=F_2(\omega_3,\omega_4).
\end{gather*}
\end{subequations}
Here $F_1$ and $F_2$ are arbitrary smooth functions of their arguments, whereby  $\omega_1$, $\omega_2$, $\omega_3$ and $\omega_4$ are given by 
(\ref{sl3-omega-1}), (\ref{sl3-omega-2}), (\ref{sl3-omega-3}) and (\ref{sl3-omega-4}), respectively.

\strut\hfill

\noindent
{\bf Example 2.2:} A system of two partial differential equations of order four that is invariant under the Möbius transformation (\ref{M-u-2-xt}) is obtained by considering, for example,
\begin{subequations}
\begin{gather*}
\omega_1=F_1(\omega_2,\omega_4,\omega_{41},\omega_{42})\\
\omega_3=F_2(\omega_2,\omega_4,\omega_{41},\omega_{42}),
\end{gather*}
\end{subequations}
where $\omega_1$, $\omega_2$, $\omega_3$, $\omega_4$, $\omega_{41}$ and $\omega_{42}$ are given by 
(\ref{sl3-omega-1}), (\ref{sl3-omega-2}), (\ref{sl3-omega-3}), (\ref{sl3-omega-4}), (\ref{old-w1}) and (\ref{old-w2}), respectively. If we drop $\omega_2$ and $\omega_4$ from the previous system and let $F_1=\omega_{41}$, $F_2=\omega_{42}$, we obtain
\begin{subequations}
\begin{gather}
\frac{u_{2,x}u_{1,xx}-u_{1,x}u_{2,xx}}{u_{2,x}u_{1,t}-u_{1,x}u_{2,t}}
=-\frac{3u_{1,x}u_{2,x}(S_{1,x}-S_{2,x})}{u_{1,x}u_{2,xx}-u_{2,x}u_{1,xx}}\nn\\[0.3cm]
\qquad\qquad\qquad\qquad\quad
-4\left(
\frac{u_{1,x}u_{2,x}}{u_{1,x}u_{2,xx}-u_{2,x}u_{1,xx}}\right)^2
(S_1-S_2)^2
\nn\\[0.3cm]
\qquad\qquad\qquad\qquad\quad
+6\left(
\frac{u_{1,x}u_{2,xx}S_1-u_{2,x}u_{1,xx}S_2}{u_{1,x}u_{2,xx}-u_{2,x}u_{1,xx}}\right)\\[0.3cm]
\frac{
u_{1,tt}u_{2,t}-u_{1,t}u_{2,tt}}
{u_{2,x}u_{1,t}-u_{1,x}u_{2,t}}
=
\frac{u_{1,x}u_{2,xx}S_{2,x}-u_{2,x}u_{1,xx}S_{1,x}}{
u_{1,x}u_{2,xx}-u_{2,x}u_{1,xx}}\nn\\[0.3cm]
\qquad\qquad\qquad\qquad\quad
+\frac{2}{3}
\left(\frac{u_{1,x}u_{2,x}}
{u_{1,x}u_{2,xx}-u_{2,x}u_{1,xx}}\right)^2
(S_{1,x}-S_{2,x})(S_1-S_2)\nn\\[0.3cm]
\qquad\qquad\qquad\qquad\quad
+\frac{2}{9}
\left(\frac{3u_{1,xx}(S_1-S_2)}{u_{1,x}u_{2,xx}-u_{2,x}u_{1,xx}}
-\frac{2u_{2,x}u_{1,x}^2(S_1-S_2)^2}{(u_{1,x}u_{2,xx}-u_{2,x}u_{1,xx})^2}\right)\times\nn\\[0.3cm]
\qquad\qquad\qquad\qquad\quad
\times\left(
3u_{2,xx}-\frac{2u_{1,x}u_{2,x}^2(S_1-S_2)}{u_{1,x}u_{2,xx}-u_{2,x}u_{1,xx}}\right),
\end{gather}
\end{subequations}
where $S_1$ and $S_2$ are the Schwarzian derivatives (\ref{Schwarzian-u1-u2})
in $u_1(x,t)$ and $u_2(x,t)$, respectively.

\section{Concluding remarks}

Given the well-known importance 
of the one-variable Möbius transformation (\ref{Mobius-u}) and its invariant, the Schwarzian derivative (\ref{Schwarzian}), for partial differential differential equations
(see Example 1.2),
 we investigate here the invariants of the two-variable Möbius transformation (\ref{M-u-2-xt}) and its relation to systems of two partial differential equations.
 Two examples of systems that are invariant under the Möbius transformation (\ref{M-u-2-xt}) are given in Example 2.1 and Example 2.2. 
It is clear from the results obtained here that there exists no system of evolution equations that is kept invariant under (\ref{M-u-2-xt}), i.e. no system of the form
\begin{gather*}
u_{1,t}=F_1(x,t,u_1.u_2,u_{1,x},u_{2,x},\ldots, u_{1,px},u_{2,qx})\\
u_{2,t}=F_2(x,t,u_1.u_2,u_{1,x},u_{2,x},\ldots, u_{1,px},u_{2,qx}).
\end{gather*}
It could be of interest to study the systems of partial differential equations that are invariant under (\ref{M-u-2-xt}) for further properties, for example to see whether there exist any 
symmetry-integrable systems amongst those, similar to the case of evolution equations for one dependent variable (see Example 1.2).  

Let us finally point to the two invariants  (\ref{old-w1}) and (\ref{old-w2}) of the Möbius transformation (\ref{M-u-n}). These 4th-order invariants consist of only $x$-derivatives of the variables
$u_1$ and $u_2$ and are, in a sense, similar to the Schwarzian derivative which is a 3rd-order invariant for (\ref{Mobius-u}). 
It remains 
to be seen whether or not (\ref{old-w1}) and (\ref{old-w2}) 
will indeed turn out to be
significant for the study of systems of differential equations and their applications.

The referee kindly pointed out the papers \cite{Beffa} and \cite{Hubert} where related aspects of differential invariants have been reported. A comparison of our results to these works could be of interest.

\label{lastpage}

\end{document}